\documentclass[aps,prd,preprint,nofootinbib,groupedaddress,showpacs,showkeys]{revtex4-1}

\usepackage{amsmath,amssymb,MnSymbol}
\usepackage{graphicx}
\usepackage[colorlinks,linkcolor=blue,anchorcolor=blue,citecolor=blue]{hyperref}

\begin{document}
\def\query#1{\marginpar{\begin{flushleft}\footnotesize#1\end{flushleft}}}
\newcommand{\br}{\bra{p',s'}}
\newcommand{\kt}{\ket{p,s}}
\newcommand{\p}{\partial_}
\newcommand{\m}{\mathbf}{}
\newcommand{\bs}{\boldsymbol}{}
\newcommand{\tf}{\tau_1}
\newcommand{\ts}{\tau_2}
\newcommand{\eq}{\begin{eqnarray}}
\newcommand{\en}{\end{eqnarray}}
\newcommand{\intsuml}{\int\hspace*{-.5cm}\sum_{~l}}
\newcommand{\intsumk}{\int\hspace*{-.5cm}\sum_{~k}}

\title{Three particle quantization condition in a finite volume:\\
2. general formalism and the analysis of data}


\author{Hans-Werner Hammer$^{a}$, Jin-Yi Pang$^{b}$ and Akaki Rusetsky$^{b}$\\~}

\affiliation{$^a$Institut f\"ur Kernphysik, Technische Universit\"at Darmstadt,
64289 Darmstadt, Germany and \\ ExtreMe Matter Institute EMMI, GSI Helmholtzzentrum f\"ur Schwerionenforschung,\\ 64291 Darmstadt, Germany\\
\\
$^b$Helmholtz-Institut f\"ur Strahlen- und Kernphysik (Theorie) and\\
Bethe Center for Theoretical Physics, Universit\"at Bonn,\\ D-53115 Bonn, Germany}

\date{\today}

\begin{abstract}
We derive the three-body quantization condition in a finite volume
using an effective field theory in the particle-dimer picture.
Moreover, we consider the extraction of physical observables from
the lattice spectrum using the quantization condition. To illustrate
the general framework, we calculate the volume-dependent three-particle 
spectrum in a simple model both below and above the three-particle 
threshold. The relation to existing approaches is discussed in detail.
\end{abstract}


\maketitle

\setcounter{section}{0}

\section{Introduction}

The extraction of physical observables in sectors containing three and more
particles remains one of the main challenges in lattice QCD. In contrast, this issue
has already been settled in the one- and two-particle sectors. Namely, in the one-particle
sector one determines the effective masses of stable particles. The infinite-volume
limit is straightforward as lattice artifacts are exponentially suppressed at large
volumes. In the case of the elastic two-body scattering, the celebrated L\"uscher 
formula~\cite{Luescher-torus} {\em algebraically}
relates the finite-volume energy eigenvalues to the infinite-volume scattering phase
shift at the same energies. The approach remains conceptually the 
same for coupled-channel inelastic scattering~\cite{Lage-KN,Lage-scalar,He,Sharpe,Briceno-multi,Liu,PengGuo-multi} 
and has already been used to analyze the data in the two-channel 
system~\cite{Wilson-pieta}. A related approach that boils down to
using a different parameterization of the infinite-volume amplitudes goes under the name
of ``unitary ChPT in a finite volume''~\cite{oset,Doring-scalar,oset-in-a-finite-volume}
and has already been used in Ref.~\cite{Bolton:2015psa} to analyze 
$P$-wave $\pi\pi$ scattering and to study the properties of the $\rho$-meson.
When the number of coupled channels is large, it might be advantageous
to directly extract the real and imaginary parts of the optical potential
in selected channels~\cite{optical} (note also the recent work~\cite{Hansen:2017mnd},
 which aims at the extraction of the {\em total} width of the
resonances decaying into the multiple channels).
 An alternative scheme aims at the extraction
of hadronic potentials from the data~\cite{HAL-essentials,HAL-derivatives}
(see, e.g., Ref.~\cite{HAL-multi} for the generalization of this approach to the 
multi-channel case).

On the one hand, there is no such framework for intermediate states with 
three or more particles, although several attempts in this direction have been 
undertaken. The first formal investigation dates back to 2012, when it was 
rigorously shown that the three-body spectrum in a finite volume is determined
solely by the three-body $S$-matrix elements in the infinite volume~\cite{Polejaeva:2012ut}.
In the following years, further important aspects of the three-body problem in
a finite volume have been addressed~\cite{Meissner:2014dea,Guo:2016fgl,Guo:2017ism,Briceno:2012rv,Hansen:2014eka,Hansen:2015zga,Hansen:2015zta,Hansen:2016fzj,Hansen:2016ync,Briceno:2017tce}.
In particular, the relativistic three-particle quantization condition in a finite volume has been obtained
in  Refs.~\cite{Hansen:2014eka,Hansen:2015zga,Briceno:2017tce}.
In their subsequent papers, the authors were able to demonstrate that the 
framework is capable of reproducing known results, e.g., for the many-body ground-state energy
or the energy shift of the three-body bound state in a finite volume. Despite this success,
the quantization condition in these papers is not yet given in a form suitable for the
analysis of the real lattice data:
The whole formalism is still very complicated and the relation to the physical
observables is not transparent. Finally, we mention
Refs.~\cite{Kreuzer:2010ti,Kreuzer:2009jp,Kreuzer:2008bi,Kreuzer:2012sr}, which
addressed the three-body problem in a finite volume numerically using an effective
field theory in the particle-dimer picture.
Their numerical results for the finite volume spectrum strongly support the statement
that the spectrum does not
depend on the off-shell behavior of the three-body amplitudes, which
was first proven in Ref.~\cite{Polejaeva:2012ut} and confirmed in Ref.~\cite{Hansen:2014eka}.
These studies also suggest a strategy for analytical investigations 
of three-body dynamics in a finite volume.

On the other hand, recent years 
have seen a steady progress in lattice simulations
involving three- and more-particle states. As a prominent example, we cite
the numerous attempts to calculate the mass of the Roper resonance
and solve the problem of the level ordering between this resonance and the 
$N^*(1535)$~\cite{Mathur:2003zf,Guadagnoli:2004wm,Leinweber:2004it,Sasaki:2005ap,Sasaki:2005ug,Burch:2006cc,Liu:2014jua,Mahbub:2010jz,Lasscock:2007ce,Lang:2016hnn}. It is a well known experimental fact that the Roper resonance decays
with a significant probability (up to 40\%) into the final state of a nucleon and two pions. Hence, a reliable extraction of its parameters is impossible without solving the three-body problem in a finite volume. Furthermore, a rapid advance
of lattice nuclear simulations~\cite{Beane:2010em,Beane:2012vq,Chang:2015qxa},
as well as chiral effective field theories on the 
lattice~\cite{Epelbaum:2009pd,Epelbaum:2011md,Rokash:2013xda,Elhatisari:2015iga},
provide us with data that can be properly analyzed, if and  only if the few-body
dynamics in a finite volume is understood. Otherwise, the extraction of
reaction rates, elastic and inelastic cross sections, etc. from
such calculations is not possible.

In our opinion, the main question which should be answered is the following:
what is the optimum set of infinite-volume parameters (observables) of the 
three-body system which can be extracted directly from the data? To illustrate
this question, we refer again to two-body elastic scattering. In this 
case, we have one measured lattice observable (the energy level) vs. one 
infinite-volume observable (the phase shift at the same energy). These two are unambiguously 
related via the L\"uscher equation. In the two-channel case, we have again one
finite volume energy level but three independent infinite-volume observables
($K$-matrix elements at the same energy). Thus, the most convenient strategy consists
of parameterizing the energy-dependence of the multi-channel 
$K$-matrix in terms of a few parameters (resonance locations, residua, 
threshold expansion parameters) and fitting all available lattice data in a given
energy interval with this parameterization 
(see, e.g.,~\cite{Wilson-pieta}). In the next step, 
having fixed these parameters,
we may reliably determine the $K$-matrix elements everywhere in the given
energy interval. 

In case of three particles, much effort has been put into obtaining an analog
of the L\"uscher equation through collecting all infinite-volume contributions
into the three-body 
$K$-matrix~\cite{Polejaeva:2012ut,Hansen:2014eka,Hansen:2015zga,Briceno:2017tce}.
The result is quite complicated and, in our opinion, is not well suited for the
analysis of lattice data. For example, ``smooth cutoffs'' should be made and an ``unconventional''
$K$-matrix should be introduced at the intermediate stage.
We argue in this paper that most of these complications stem from the inappropriate
choice of parameters. Using the particle-dimer formalism,
we arrive at a rather simple parameterization of the infinite-volume
three-body $S$-matrix as well as the three-body spectrum in a finite volume.
This provides a framework for the analysis of lattice data.

The layout of the paper is as follows.
In Section~\ref{sec:infinite}, we briefly review the 
infinite-volume framework for the description of the two- and three-body 
sectors, which is based on non-relativistic effective Lagrangians.
The transition to the particle-dimer picture and the issue of the off-shell
behavior is discussed in detail. In Section~\ref{sec:finite}, we consider 
the same theory in a finite volume and discuss the strategy for the lattice
data analysis. A simple illustration is provided for the statement that
the finite-volume spectrum is determined only by on-shell three-body
$S$-matrix elements. In this section, we also present the results of the
numerical calculations of the volume-dependent three-body spectrum, both
below and above the three-particle threshold. In Section~\ref{sec:comparison}, we
make a detailed comparison with the existing approaches. Finally, 
Section~\ref{sec:concl} contains our conclusions.

\section{Infinite volume}
\label{sec:infinite}

\subsection{Two-particle sector}

In order to simplify the formalism and highlight the 
central conceptual issues, we consider the interaction of three identical 
non-relativistic scalars. In addition, we assume that two-to-three 
particle transitions are forbidden. Thus for applications to QCD,
our final result still needs to be decorated with spin indices,
relativistic boosts into non-rest frames, etc. These  effects can
be included in a second stage. For example, the coupling of
the two- and three-particle sectors can be included along the lines
of Ref.~\cite{Briceno:2017tce}. We stress that none of these 
issues affect the essence of the problem considered in this paper.

The three-particle Lagrangian can be written in the following form
\eq
{\cal L}=\psi^\dagger\biggl(i\partial_0+\frac{\nabla^2}{2m}\biggr)\psi
+{\cal L}_2+{\cal L}_3\, ,
\en
where $\psi(x)$ denotes a non-relativistic field with the propagator
\eq
i\langle 0|T\psi(x)\psi^\dagger(y)|0\rangle
=\int\frac{d^4p}{(2\pi)^4}\,
\frac{e^{-ip(x-y)}}{w({\bf p})-p_0-i0}\, ,\quad\quad
w({\bf p})=\frac{{\bf p}^2}{2m}\, ,
\en
$m$ is the mass of the particle, and ${\cal L}_2,{\cal L}_3$ denote the two-
and three-particle interaction terms, respectively.
 
Let us start from the two-particle term. It contains a tower of operators
of increasing mass dimension or, equivalently, an increasing number of
space derivatives
\eq
{\cal L}_2=-\frac{C_0}{2}\,\psi^\dagger\psi^\dagger\psi\psi
+\frac{C_2}{4}\,(\psi^\dagger\tensor{\nabla}^2\psi^\dagger\psi\psi+\mbox{h.c.})
+O(\nabla^4)\, .
\label{eq:L2C0C2}
\en
Here, the first term corresponds to a non-derivative interaction which is
purely S-wave. In the second term, we employ the Galilean invariant derivative operator
$\tensor{\nabla} \equiv (\roarrow{\nabla} - \loarrow{\nabla})/2$, which is understood
to act only on the fields immediately left and right of the operator.
For identical spinless particles there are no
odd partial waves, so the contribution of higher partial waves starts
at $O(\nabla^4)$.

It is more convenient to use the momentum representation for the analysis
of the higher-order terms.
Because of Galilei invariance, the interaction does not depend on the center-of-mass
momentum. It is characterized by
the relative momenta of the two particles in the final and initial states,
${\bf p}$ and ${\bf q}$, respectively. Using rotational invariance and Bose-symmetry,
the matrix element of the interaction Lagrangian between the two-particle states
can be written in the form
\eq
\langle{\bf P}, {\bf p}|{\cal L}_2|{\bf q}, {\bf P}\rangle
=\sum_{n,m,k=0}^\infty l_{nmk}{\bf p}^{2n}{\bf q}^{2m}({\bf p}{\bf q})^{2k}\, ,
\en
where $l_{nmk}$ are linear combinations of $C_0,C_2,\ldots$.

Furthermore, the expression $q_{i_1}\cdots q_{i_{2k}}$ is given by a sum of a traceless
tensors of rank $2k$ and less, and terms containing the Kronecker symbol. For example,
\eq
q_iq_j=\biggl(q_iq_j-\frac{1}{3}\,\delta_{ij}{\bf q}^2\biggr)+ \frac{1}{3}\,\delta_{ij}{\bf q}^2\,,
\en
and similar expressions hold for higher order tensors. The first term in brackets is traceless
and corresponds to a D-wave. The Kronecker symbol $\delta_{ij}$ in the second term will be convoluted
with $p_ip_j$ and yields ${\bf p}^2$, corresponding to a S-wave. Continuing this way, we obtain
\eq\label{eq:TT}
\langle {\bf P}, {\bf p}|{\cal L}_2|{\bf q}, {\bf P}\rangle
=\sum_{n,m,k=0}^\infty l'_{nmk}{\bf p}^{2n}{\bf q}^{2m}
\sum_{i_1,\cdots i_{2k}} F_{i_1,\cdots i_{2k}}({\bf p}) F_{i_1,\cdots i_{2k}}({\bf q})\, ,
\en
where the $F_{i_1,\cdots i_{2k}}$ are traceless tensors in all indices and
correspond to even orbital momentum $L=2k$, whereas $l'_{nmk}$
are linear combinations of $l_{nmk}$. Hence, we obtain a clear-cut classification
of the operators of the Lagrangian in the partial waves and -- within a sector
with a given value of the orbital momentum -- in powers of momenta 
${\bf p}^{2n}{\bf q}^{2m}$.

Next, we consider the matching of the couplings $C_0,C_2,\ldots$ to the physical
observables. We limit ourselves to a sector with a fixed value of the
orbital momentum (say, the S-wave). The available observables for fitting
are the effective-range parameters\footnote{Usually, the effective-range
expansion is performed around threshold. This limits the applicability of the method
to small values of the three-momenta. However, the quantity $p\cot\delta(p)$ can also
be expanded in powers of $p^2$ near some $p^2=p_0^2$ instead of $p^2=0$.
This corresponds to a rearrangement of the perturbation series without changing the total
result (the Lagrangian that leads to the modified series can easily be written down).
It is expected that with an appropriate choice of $p_0^2$, the convergence of the series
in a limited interval of momenta can be improved.}
\eq\label{eq:ERE}
p\cot\delta(p)=-\frac{1}{a}+\frac{r_e}{2}\,p^2+\sum_{n=2}^\infty b_{2n}p^{2n}\, ,
\quad\quad p^2={\bf p}^2\, .
\en
At order ${\bf p}^4$ we have two independent operators
\eq\label{eq:L2}
{\cal L}_2^{S-wave}&=&-\frac{C_0}{2}\,\psi^\dagger\psi^\dagger\psi\psi
+\frac{C_2}{4}\,(\psi^\dagger\tensor{\nabla}^2\psi^\dagger\psi\psi+\mbox{h.c.})
\nonumber\\[2mm]
&-&\frac{C_4}{4}\,(\psi^\dagger\tensor{\nabla}^4\psi^\dagger\psi\psi+\psi^\dagger\tensor{\nabla}^2\psi^\dagger\psi\tensor{\nabla}^2\psi+\mbox{h.c.})
\nonumber\\[2mm]
&-&\frac{C_4'}{4}\,(\psi^\dagger\tensor{\nabla}^4\psi^\dagger\psi\psi-\psi^\dagger\tensor{\nabla}^2\psi^\dagger\psi\tensor{\nabla}^2\psi+\mbox{h.c.})
+O(\nabla^6)\, ,
\nonumber\\[2mm]
\langle {\bf P}, {\bf p}|{\cal L}_2^{S-wave}|{\bf q},{\bf P}\rangle
&=&-2C_0-C_2({\bf p}^2+{\bf q}^2)-C_4({\bf p}^2+{\bf q}^2)^2
-C_4'({\bf p}^2-{\bf q}^2)^2\, .
\en
It is clear that the matching to $b_4$ from Eq.~(\ref{eq:ERE}) determines the
constant $C_4$, whereas the term with $C_4'$ vanishes on the energy shell
${\bf p}^2={\bf q}^2$ and thus is not fixed from matching. Moreover, using
the equations of motion
\eq
\biggl(i\partial_0+\frac{\nabla^2}{2m}\biggr)\psi=C_0\psi^\dagger\psi\psi+\cdots\, ,
\en
we find that the off-shell term is proportional to a total time derivative (modulo surface terms)
\eq
\psi^\dagger\tensor{\nabla}^4\psi^\dagger\psi\psi-\psi^\dagger\tensor{\nabla}^2\psi^\dagger\psi\tensor{\nabla}^2\psi+\mbox{h.c.}\propto
\partial_0^2(\psi^\dagger\psi^\dagger\psi\psi)\, ,
\en
and, hence, does not affect the equations of motion. Note that the above relation holds up to terms containing
six fields. Consequently, if the three-particle sector is also considered, the off-shell terms in the two-particle sector
can be eliminated in favor of the three-particle forces.

Another statement concerns the two-particle spectrum in a finite volume.
The validity of the L\"uscher equation implies that such off-shell terms do not
affect the spectrum, which is solely determined by the on-shell 
$S$-matrix elements. Physically, this stems from the existence of two widely
separated scales -- the box size $L$ and the typical interaction range $R$,
with $R\ll L$. This means that the two-particle wave function near the boundaries
is given by its asymptotic form determined by the phase shift. Hence, only this
quantity enters the finite-volume quantization condition. Our aim is to 
verify the same statement in the three-particle sector as well, where it looks less intuitive.
Namely, there exist regions in the configuration space, where two out 
of three-particles are
close to each other and the third particle is far away (at the distances of order of the box size $L$).
Nevertheless, as we shall see, the statement still holds.

\begin{figure}[t]
\begin{center}
\includegraphics*[width=10.cm]{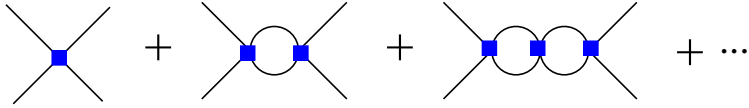}
\caption{$T$-matrix in the non-relativistic theory. The filled boxes
represent the tree-level Lagrangian containing the low-energy couplings $C_0,C_2,C_4,\ldots$.}
\label{fig:bubbles}
\end{center}
\end{figure}

The two-body scattering $T$-matrix is given by the sum of the bubble diagrams
shown in Fig.~\ref{fig:bubbles}. On the mass shell, is is equal to
\eq\label{eq:T}
T=\frac{8\pi/m}{p\cot\delta(p)-ip}\, ,
\en
where $p\cot\delta(p)$ is given by the effective-range expansion~(\ref{eq:ERE}).


\subsection{Dimer formalism}

Next, we consider the introduction of the dimer field in the context of the
pure two-body problem. Note first that it is allowed to rewrite the above
expression in a form
\eq\label{eq:TThatf}
\langle {\bf P},{\bf q}|{\cal L}_2|{\bf p},{\bf P}\rangle
=\sigma\sum_{k=0}^\infty f({\bf p}^2)f({\bf q}^2)
\sum_{i_1,\cdots i_{2k}} F_{i_1,\cdots i_{2k}}({\bf p}) F_{i_1,\cdots i_{2k}}({\bf q})\, ,
\en
where $f({\bf p}^2)=f_0+f_1{\bf p}^2+\ldots$.
Further, $\sigma=-1$ if $C_0>0$ and {\em vice versa.} 
Since, at a given order in
${\bf p}^2$, there is only one physically relevant constant that can be matched
on shell, one may recursively express the couplings $f_0,f_1,\ldots$ through
the effective range parameters.  

Now, let $T_{i_1,\cdots i_{2k}}$ 
be a field completely traceless in its indices, which describes a dimer with
spin equal to $2k$. One may write down the Lagrangian
\eq\label{eq:Ldim2}
{\cal L}&=&\psi^\dagger\biggl(i\partial_0+\frac{\nabla^2}{2m}\biggr)\psi
+\sigma\sum_{k=0}^\infty\sum_{i_1,\cdots i_{2k}} T_{i_1,\cdots i_{2k}}^\dagger T_{i_1,\cdots i_{2k}}
\nonumber\\[2mm]
&+&
\biggl(\sum_{k=0}^\infty\sum_{i_1,\cdots i_{2k}}T_{i_1,\cdots i_{2k}}^\dagger
\psi\bigl[ f(-i\nabla) F_{i_1,\cdots i_{2k}}(-i\nabla)\bigr]\psi+\mbox{h.c.}\biggr)\, ,
\en
where $\psi\bigl[f(-i\nabla) F_{i_1,\cdots i_{2k}}(-i\nabla)\bigr]\psi$
is a shorthand notation (the operator $\nabla$ in the second line should be interpreted as a 
``Galilei-invariant'' operator, see the discussion below Eq.~(\ref{eq:L2C0C2})).
To explain this notation, consider first the 
spinless dimer $k=0$. Writing down $f(-i\nabla)=f_0+f_1(-i\nabla)^2+\ldots$, 
and taking into account the Bose-symmetry factors, we get
\eq
\psi\bigl[f_0\bigr]\psi&=&\frac{1}{2}\,f_0\psi\psi\, ,
\nonumber\\[2mm]
\psi\bigl[f_1(-i\nabla)^2\bigr]\psi&=&-\frac{1}{4}\,f_1
(\psi\nabla^2\psi-\nabla_i\psi\nabla_i\psi)\, ,
\en
and so on. For the spin-2 dimer (i.e., $k=1$ we have)
\eq
\psi\bigl[F_{ij}(-i\nabla)\bigr]\psi
&=&\psi\biggl[\frac{3}{2}(-i\nabla)_i(-i\nabla)_j-\frac{1}{2}\,\delta_{ij}(-i\nabla)^2\biggr]\psi
\nonumber\\[2mm]
&=&-\frac{3}{8}\,\psi\nabla_i\nabla_j\psi+\frac{3}{8}\,\nabla_i\psi\nabla_j\psi
+\frac{1}{8}\,\delta_{ij}\psi\nabla^2\psi
-\frac{1}{8}\,\delta_{ij}\nabla_k\psi\nabla_k\psi\, ,
\en
and so on. In the CM frame, one may perform the partial integration, leading to the same result as before.

The dimer field $T_{i_1,\cdots i_{2k}}$ is an auxiliary field. Integrating it out
with the use of the equations of motion, or using the path integral,
it is straightforward to verify that the dimer formalism on the energy shell
is mathematically
equivalent to the theory defined by the two-body Lagrangian, Eq.~(\ref{eq:L2}). 
In particular, the on-shell 
$T$-matrix is again given by Eq.~(\ref{eq:T}), also in non-rest frames.
Putting it differently, the dimer picture is not an approximation -- it is
an alternative description of two-body scattering. The scattering in higher partial waves is represented through
dimers with arbitrary integer spins.
 
\subsection{Three-particle sector}

Next, let us turn to the three-particle sector and write down the Lagrangian. At leading
order, it is given by
\eq
{\cal L}_3^{LO}=-\frac{D_0}{6}\,\psi^\dagger\psi^\dagger\psi^\dagger\psi\psi\psi\, .
\en
Further, we turn to the next-to-leading order.
As in the two-particle sector, it is easier to carry out the classification
of the derivative terms in momentum space.
To keep the discussion transparent, we restrict ourselves to the center-of-mass frame
${\bf p}_1+{\bf p}_2+{\bf p}_3={\bf q}_1+{\bf q}_2+{\bf q}_3={\bf 0}$
in the three-particle sector. It is always possible to rewrite the
expressions in a manifestly Galilei-invariant form using Galilei invariant
derivatives.
At next-to-leading order, we have the following invariants:
\eq
{\bf p}_i^2\,, ~ {\bf p}_i{\bf p}_j\,,~
{\bf q}_i^2\,, ~ {\bf q}_i{\bf q}_j\,,~
{\bf p}_i{\bf q}_i\,, ~ {\bf p}_i{\bf q}_j\,,\quad\quad
i\neq j\, ,\quad i,j=1,2,3\, .
\en
Bose-symmetry, time invariance and Hermiticity exclude all structures
but one
\eq
\langle {\bf q}|{\cal L}_3^{NLO}|{\bf p}\rangle=
D_2\sum_{i=1}^3({\bf p}_i^2+{\bf q}_i^2)
\en
which in position space gives
\eq
{\cal L}_3^{NLO}=-\frac{D_2}{12}\,(\psi^\dagger\psi^\dagger\nabla^2\psi^\dagger\,
 \psi\psi\psi
+\mbox{h.c.})\, .
\en
Let us now consider next-to-next-to-leading order.
Using the condition 
${\bf p}_1+{\bf p}_2+{\bf p}_3={\bf q}_1+{\bf q}_2+{\bf q}_3={\bf 0}$ and
taking into account the Hermiticity of the Lagrangian and Bose-symmetry, 
the set of linearly independent invariants is
\eq
O&=&({\bf p}_1^4+{\bf p}_2^4+{\bf p}_3^4)+({\bf q}_1^4+{\bf q}_2^4+{\bf q}_3^4)\, ,
\nonumber\\[2mm]
O'&=&\bigl[({\bf p}_1^2+{\bf p}_2^2+{\bf p}_3^2)+({\bf q}_1^2+{\bf q}_2^2+{\bf q}_3^2)\bigr]^2\, ,
\nonumber\\[2mm]
O''&=&\bigl[({\bf p}_1^2+{\bf p}_2^2+{\bf p}_3^2)-({\bf q}_1^2+{\bf q}_2^2+{\bf q}_3^2)\bigr]^2\, .
\en
Note that we have written down only the terms that contribute to the S-wave.
There are also D-wave contributions generated at this order, say, by
terms of the type $\sum\limits_{i,j=1}^3({\bf p}_i{\bf q}_j)^2$.
For illustrative purposes, however, we restrict ourselves to the S-wave 
contribution only.

In position space, we get
\eq
{\cal L}_3^{NNLO}&=&\frac{D_4}{12}\,\bigl(\psi^\dagger\psi^\dagger\nabla^4\psi^\dagger\,\psi\psi\psi+\mbox{h.c.}\bigr)
\nonumber\\[2mm]
&+&\frac{D_4'}{12}\,\bigl((\psi^\dagger\psi^\dagger\nabla^4\psi^\dagger\,\psi\psi\psi+2\psi^\dagger\nabla^2\psi^\dagger\nabla^2\psi^\dagger\,\psi\psi\psi+\mbox{h.c.})+6\psi^\dagger\psi^\dagger\nabla^2\psi^\dagger\,\psi\psi\nabla^2\psi\bigr)
\nonumber\\[2mm]
&+&\frac{D_4''}{12}\,\bigl((\psi^\dagger\psi^\dagger\nabla^4\psi^\dagger\,\psi\psi\psi+2\psi^\dagger\nabla^2\psi^\dagger\nabla^2\psi^\dagger\,\psi\psi\psi+\mbox{h.c.})-6\psi^\dagger\psi^\dagger\nabla^2\psi^\dagger\,\psi\psi\nabla^2\psi\bigr)
\en
The low-energy coupling $D_4''$ is analogous to the ``off-shell''
coupling $C_4'$ considered
in the two-particle sector -- at tree level, it does not contribute to
the three-particle amplitude.

Furthermore, using the equations of motion it can be shown that
\eq
-(2m)^2\partial_0^2(\psi^\dagger\psi^\dagger\psi^\dagger)&=&
6\psi^\dagger\nabla^2\psi^\dagger\nabla^2\psi^\dagger
+3\psi^\dagger\psi^\dagger\nabla^4\psi^\dagger+\cdots\,
\nonumber\\[2mm]
-(2m)^2\partial_0^2(\psi\psi\psi)&=&
6\psi\nabla^2\psi\nabla^2\psi
+3\psi\psi\nabla^4\psi+\cdots\, ,
\nonumber\\[2mm]
-(2m)^2\partial_0(\psi^\dagger\psi^\dagger\psi^\dagger)\partial_0(\psi\psi\psi)&=&
-9\psi^\dagger\psi^\dagger\nabla^2\psi^\dagger\psi\psi\nabla^2\psi+\cdots\, ,
\en
where the ellipses stand for terms containing more field operators.
Taking these equations into account, it is straightforward to see that
the term proportional to $D_4''$ is a total time derivative
\eq
-\frac{D_4''}{36}\,(2m)^2
\partial_0^2(\psi^\dagger\psi^\dagger\psi^\dagger\psi\psi\psi)\,,
\en
and therefore does not contribute to the equations of motion.

\subsection{Insertion of the off-shell terms into Feynman diagrams}

As we have seen, the on-shell three-body scattering $T$-matrix does not
depend on the low-energy coupling $D_4''$. We now want to expand this argument
beyond tree level and show that the $T$-matrix does not depend on $D_4''$ 
at all. This would have been easy in the two-body sector: one would use
dimensional regularization and argue that the off-shell terms in
 perturbation theory lead to 
no-scale integrals that vanish in this regularization. The final answer then 
should not depend on the regularization used. This argument, however, rests
on the fact that the two-body potential is a low-energy polynomial. This
is not true anymore in the three-body sector, where the pair interactions give rise
to a non-polynomial potential. For this reason, we have
to examine the perturbation series in the three-body sector more carefully.

The three-body $T$-matrix obeys the Lippmann-Schwinger equation
\eq
T=V+VG_0T\, ,
\en
where $G_0$ is the free three-body Green function, and $V$ denotes the kernel
(potential) of this equation -- the sum of all Feynman diagrams which can not
be made disconnected by cutting exactly three particle lines. Of course, it
is well known that the above equation is mathematically ill-defined. All 
two-body interactions should be summed up first, leading to the Faddeev 
equations. This fact, however, will not bother us in the following, since
we consider the Lippmann-Schwinger equation merely as a tool to generate a
full set of Feynman diagrams in the three-body $T$-matrix.

According to the discussion in the previous subsection, we may split off
the off-shell part from the potential $V$:
\eq
V=\bar V+V'\, ,
\en
where, in our case, $V'=\kappa(E_f-E_i)^2,~\kappa\propto D_4''$. Here, 
$E_i=\frac{1}{2m}\,({\bf q}_1^2+{\bf q}_2^2+{\bf q}_3^2)$ and
$E_f=\frac{1}{2m}\,({\bf p}_1^2+{\bf p}_2^2+{\bf p}_3^2)$ are the
initial and final three-particle energies, respectively. Further, it can be straightforwardly
shown that
\eq
T=T'+(1+T'G_0)\nu(1+G_0T')\, ,
\en
where $T'$ is the scattering matrix of the ``off-shell'' potential $V'$ only:
\eq
T'=V'+V'G_0T'
\en
and $\nu$ obeys the following equation
\eq\label{eq:bartau}
\nu=\bar V+\bar V(G_0+G_0T'G_0)\nu=\bar\nu+\bar\nu(G_0T'G_0)\nu\, ,
\en 
where $\bar\nu$ is the scattering matrix on the potential $\bar V$ (without the ``off-shell'' term):
\eq\label{eq:bartau_def}
\bar\nu=\bar V+\bar VG_0\bar\nu\, .
\en
Our aim is to demonstrate that
\eq\label{eq:aim}
T\biggr|_{\mbox{on shell}}=\bar\nu\biggr|_{\mbox{on shell}}\, .
\en
This aim can be achieved in a few consecutive steps. First of all, it is 
immediately seen that $T'$ vanishes on shell. Indeed,
using dimensional regularization, we get
\eq\label{eq:T1z}
T'(E)=\kappa(E_f-E_i)^2+\kappa^2(E_f-E)^2g_0(E)(E-E_i)^2\, .
\en
Here, $E$ is the off-shell energy (in general, $E\neq E_i\neq E_f$), and
\eq
g_0(E)=\int_{\bf k}\frac{1}{E({\bf k})-E-i0}\, , 
\en
where
\eq
\int_{\bf k}\doteq \int
\frac{d^d{\bf k}_1}{(2\pi)^3}\,\frac{d^d{\bf k}_2}{(2\pi)^3}\,\frac{d^d{\bf k}_3}{(2\pi)^3}\,(2\pi)^d\delta^d({\bf k}_1+{\bf k}_2+{\bf k}_3)\, ,\quad\quad
E({\bf k})=\frac{1}{2m}\,({\bf k}_1^2+{\bf k}_2^2+{\bf k}_3^2)\, .
\en
The expression (\ref{eq:T1z}) vanishes at $E_i=E_f=E$.

\begin{figure}[t]
\begin{center}
\includegraphics*[width=14cm]{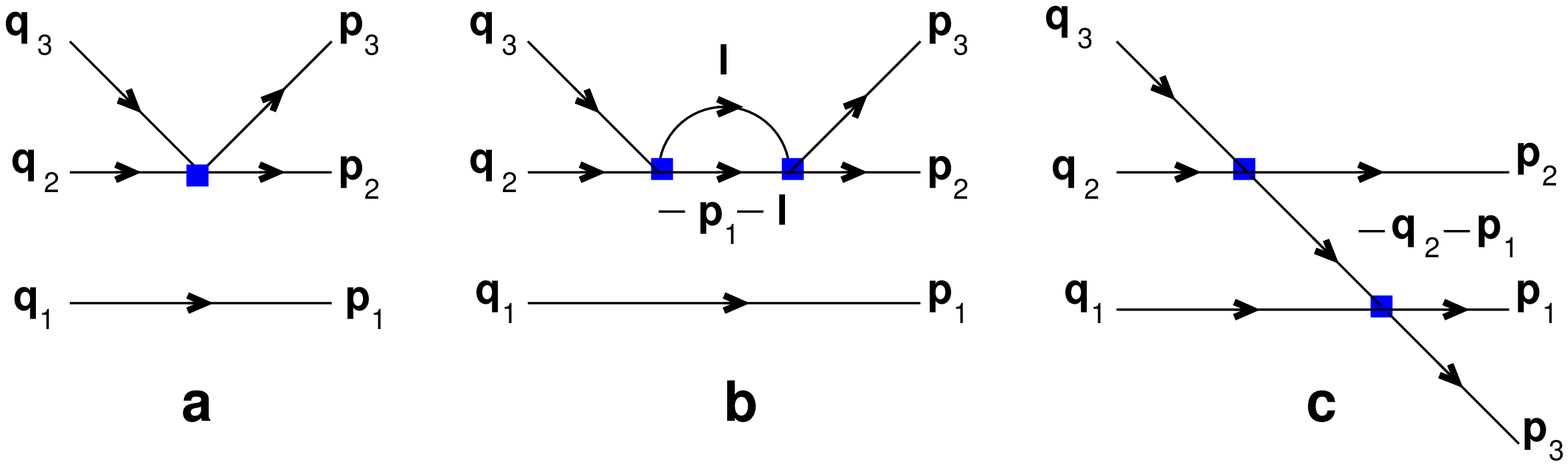}
\caption{Lowest-order Feynman graphs emerging in the expansion of the quantity $\nu$.}
\label{fig:typical}
\end{center}
\end{figure}

Further, let us consider the quantity $T'G_0\nu$ on shell:
\eq
T'G_0\nu=\int_{\bf k}(\kappa(E_f-E_{\bf k})^2+\kappa^2(E_f-E)^2g_0(E)(E-E_{\bf k})^2)
\frac{1}{E_{\bf k}-E-i0}\,\nu\, .
\en
The second term vanishes at $E_f=E$, while the first term can on  the energy shell be rewritten
as
\eq\label{eq:onshell}
T'G_0\nu\biggr|_{\mbox{on shell}}=\kappa\int_{\bf k}(E-E_{\bf k})\nu\, .
\en
This expression also vanishes, if $\nu$ is a low-energy polynomial. However, in general
it is not. For this reason, let us consider a few typical diagrams shown in Fig.~\ref{fig:typical},
in order to ensure that the above expression still leads to no-scale integrals.
Let us start from the diagram shown in Fig.~\ref{fig:typical}a which yields
$\nu=2C_0(2\pi)^d\delta^d({\bf k_1}-{\bf p}_1)$. Substituting this expression into 
into Eq.~(\ref{eq:onshell}), we get
\eq
T'G_0\nu\biggr|_{\mbox{on shell}}=2C_0\kappa\int\frac{d^d{\bf k}_2}{(2\pi)^d}\,
\biggl(E-\frac{1}{2m}\bigl({\bf p}_1^2+{\bf k}_2^2+({\bf k}_2+{\bf p}_1)^2\bigr)\biggr)=0\, .
\en
Next, let us consider the one-loop diagram shown in Fig.~\ref{fig:typical}b. Carrying out
the same steps as above, one obtains
\eq
T'G_0\nu\biggr|_{\mbox{on shell}}&=&4C_0^2\kappa\int\frac{d^d{\bf k}_2}{(2\pi)^d}\,
\frac{d^d{\bf l}}{(2\pi)^d}\,
\biggl(E-\frac{1}{2m}\bigl({\bf p}_1^2+{\bf k}_2^2+({\bf k}_2+{\bf p}_1)^2\bigr)\biggr)
\nonumber\\[2mm]
&\times&\frac{2m}{{\bf p}_1^2+{\bf l}^2+({\bf p}_1+{\bf l})^2-2mE-i0}=0\, .
\en
The last equality stems from the fact that the integral over ${\bf k}_2$ is a no-scale
integral, even if $\nu$ is not a low-energy polynomial.

Finally, we consider the diagram shown in Fig.~\ref{fig:typical}c. The result is
given by
\eq
T'G_0\nu\biggr|_{\mbox{on shell}}&=&4C_0^2\kappa\int\frac{d^d{\bf k}_1}{(2\pi)^d}\,
\frac{d^d{\bf k}_2}{(2\pi)^d}\,
\biggl(E-\frac{1}{2m}\bigl({\bf k}_1^2+{\bf k}_2^2+({\bf k}_1+{\bf k}_2)^2\bigr)\biggr)
\nonumber\\[2mm]
&\times&\frac{2m}{{\bf p}_2^2+{\bf k}_1^2+({\bf p}_2+{\bf k}_1)^2-2mE-i0}=0\, .
\en
This integral also vanishes because it is a no-scale integral with respect to the momentum ${\bf k}_2$.

This exhausts the list of the possible alternatives. Consequently, in general
$T'G_0\nu$ vanishes on shell, even if $\nu$ is not a low-energy polynomial.
By the same token, $\nu G_0T'$ and $T'G_0\nu G_0T'$ vanish on shell as well,
leading to the conclusion that $T$ and $\nu$ coincide on the energy shell.

Finally, we focus of Eq.~(\ref{eq:bartau}), which relates the quantities $\nu$ and $\bar\nu$.
Consider the second iteration of this equation, $\bar\nu G_0T'G_0\bar\nu$, where
$T'$ is given by Eq.~(\ref{eq:T1z}). It is immediately seen that the contribution 
from the second term in  Eq.~(\ref{eq:T1z})
factorizes and vanishes on the energy shell, leading to an integral similar to 
the one given in Eq.~(\ref{eq:onshell}). Using the simple algebraic identity
\eq
\frac{(E_f-E_i)^2}{(E_f-E)(E_i-E)}=\frac{E_f-E_i}{E_i-E}-\frac{E_f-E_i}{E_f-E}\, ,
\en
the contribution from the first term can be also reduced to the integrals of the same 
type. Higher order terms can be considered in a similar way. Thus, on shell $\nu=\bar\nu$ and,
consequently, Eq.~(\ref{eq:aim}) holds.

To summarize this rather lengthy but straightforward discussion, 
note that $\bar\nu$, defined by Eq.~(\ref{eq:bartau_def}) is the three-body scattering
$T$-matrix for $D_4''=0$. Thus, we have shown
that the coupling constant $D_4''$ does not appear in the on-shell $T$ matrix 
(neither at tree level, nor as an insertion in higher-order Feynman diagrams)
and can not
be fixed from experimental input. We remind the reader that this coupling multiplies
an operator that can be reduced to a total time derivative by using the equation 
of motion. In the infinite volume, the omission of such operators can not lead
to observable consequences.

\subsection{The dimer formalism in the three-particle sector}

In this subsection, we write down the particle-dimer interaction Lagrangian, which
is equivalent to the original three-particle Lagrangian. For simplicity, we restrict ourselves
to the case of a scalar dimer. The generalization to the higher partial waves is straightforward. 

We start from the Lagrangian (cf.~Eq.~(\ref{eq:Ldim2}) ),
\eq
{\cal L}&=&\psi^\dagger\biggl(i\partial_0+\frac{\nabla^2}{2m}\biggr)\psi
+\sigma T^\dagger T
+
\biggl(T^\dagger
\psi\bigl[ f(-i\nabla)\bigr]\psi+\mbox{h.c.}\biggr)
\nonumber\\[2mm]
&+&h_0T^\dagger T\psi^\dagger\psi
+h_2T^\dagger T(\psi^\dagger\nabla^2\psi+\nabla^2\psi^\dagger\psi)
\nonumber\\[2mm]
&+&h_4T^\dagger T(\psi^\dagger\nabla^4\psi+\nabla^4\psi^\dagger\psi)
+h_4'T^\dagger T \nabla^2\psi^\dagger\nabla^2\psi+\cdots\, ,
\en
where
\eq
\psi\bigl[ f(-i\nabla)\bigr]\psi&=&\frac{1}{2}\,\biggl(f_0\psi\psi+f_1\psi(-i\stackrel{\leftrightarrow}{\nabla})^2\psi
+f_2\psi(-i\stackrel{\leftrightarrow}{\nabla})^4\psi+\cdots\biggr)\, ,
\nonumber\\[2mm]
\psi(-i\stackrel{\leftrightarrow}{\nabla})^2\psi&=&\frac{1}{2}\,\biggl(-\psi\nabla^2\psi+\nabla_i\psi\nabla_i\psi\biggr)\, ,\quad\mbox{and so on.}
\en
Integrating out the dimer field, this Lagrangian can be rewritten as
\eq
{\cal L}&=&\psi^\dagger\biggl(i\partial_0+\frac{\nabla^2}{2m}\biggr)\psi
\nonumber\\[2mm]
&-&\frac{(\psi\bigl[ f(-i\nabla)\bigr]\psi)^\dagger(\psi\bigl[ f(-i\nabla)\bigr]\psi)}
{\sigma+h_0\psi^\dagger\psi
+h_2(\psi^\dagger\nabla^2\psi+\nabla^2\psi^\dagger\psi)
+h_4(\psi^\dagger\nabla^4\psi+\nabla^4\psi^\dagger\psi)
+h_4' \nabla^2\psi^\dagger\nabla^2\psi}+\cdots\,.
\en
Expanding the denominator, one gets the previous result in the two-particle sector.
In the three-particle sector, the following operators emerge:
\eq
{\cal L}_3=\frac{f_0^2h_0}{4}\,\psi^\dagger\psi^\dagger\psi^\dagger\psi\psi\psi
+\biggl(-\frac{3f_0f_1h_0}{16}+\frac{f_0^2h_2}{4} \biggr)(\psi^\dagger\psi^\dagger\nabla^2\psi^\dagger\psi\psi\psi+\mbox{h.c.})
+O(\nabla^4)\, .
\en
Here, we have used the fact that in the three-particle CM frame,
\eq
(\psi^\dagger\nabla^2\psi^\dagger-\nabla_i\psi^\dagger\nabla_i\psi^\dagger)\psi^\dagger\psi\psi\psi=\biggl(\psi^\dagger\psi^\dagger\nabla^2\psi^\dagger
-\frac{1}{2}\,\nabla_i(\psi^\dagger\psi^\dagger)\nabla_i\psi^\dagger\biggr)\psi\psi\psi
=\frac{3}{2}\,\psi^\dagger\psi^\dagger\nabla^2\psi^\dagger\psi\psi\psi\, .
\nonumber\\
\en
Consequently,
\begin{itemize}
\item
At order $\nabla^0$, the coupling constant $D_0$ is matched to $h_0$.
\item
At order $\nabla^2$, the coupling constant $D_2$ is matched to $h_2$.
\item
At order $\nabla^4$ we have three coupling constants $D_4,D_4',D_4''$ in the three-body formalism, to be matched to the two constants $h_4,h_4'$ in the particle-dimer formalism. This shows once more that one of the couplings ($D_4''$) is redundant and can be eliminated from the theory without changing the physical content.
\end{itemize}

Note also that the off-shell terms in particle-dimer scattering are essential and should
not be eliminated. This is because the dimer does not have a fixed mass. The on-shell
three-particle $T$-matrix can uniquely be related to the off-shell particle-dimer $T$-matrix.

\subsection{The scattering equation}
\label{sec:scatt_eq}

\begin{figure}[t]
\begin{center}
\includegraphics*[width=8.cm]{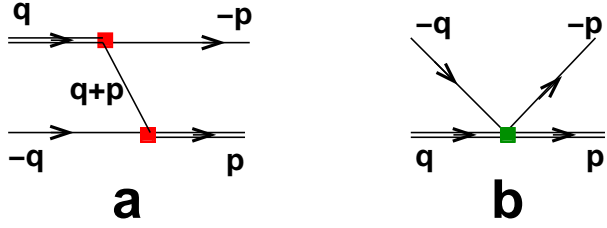}
\caption{The kernel of the three-particle Lippmann-Schwinger equation:
a) the exchange diagram between the particle and the dimer;
b) local particle-dimer interactions with couplings $h_0,h_2,\ldots$.}
\label{fig:BSE}
\end{center}
\end{figure}

In the simplest case, when only non-derivative interactions are present in the Lagrangian,
the particle-dimer scattering equation takes the form (see Fig.~\ref{fig:BSE})
\eq\label{eq:BSA}
{\cal M}({\bf p},{\bf q};E)=Z({\bf p},{\bf q};E)
+8\pi\int^\Lambda\frac{d^3{\bf k}}{(2\pi)^3}\,Z({\bf p},{\bf k};E)\tau({\bf k};E)
{\cal M}({\bf k},{\bf q};E)\, .
\en
Here, $E$ denotes the total CM energy of three particles, $a$ denotes the S-wave
scattering length, and
\eq
Z({\bf p},{\bf q};E)&=&\frac{1}{-mE+{\bf p}^2+{\bf q}^2+{\bf p}{\bf q}}
+\frac{h_0}{mf_0^2}\, ,
\nonumber\\[2mm]
\tau({\bf k};E)&=&\frac{1}{-a^{-1}+\sqrt{\frac{3}{4}\,{\bf k}^2-mE}}\, .
\en
The solution of the above equation unique because of the presence of
the ultraviolet cutoff $\Lambda$. Then, in order to ensure the independence of the 
observables on the cutoff, as $\Lambda\to\infty$, the three-body coupling
$H_0(\Lambda)=\Lambda^2h_0(\Lambda)/(mf_0^2)$ should depend on $\Lambda$
in a log-periodic manner~\cite{Bedaque:1998kg,Bedaque:1998km}.

The inclusion of the higher-order terms (still in the S-wave)
boils down to two modifications:
\begin{itemize}
\item[a)]
Replacing $\tau({\bf q};E)$ with the exact propagator
\eq
\tau({\bf q};E)&=&\frac{1}{p^*\cot\delta(p^*)+\sqrt{\frac{3}{4}\,{\bf q}^2-mE}}\, ,
\quad\quad
(p^*)^2=mE-\frac{3}{4}\,{\bf q}^2\, .
\en
\item[b)]
Adding derivative particle-dimer couplings to $Z$:
\eq
H_0(\Lambda)\to H_0(\Lambda)+H_2(\Lambda)({\bf p}^2+{\bf q}^2)+\cdots\, .
\en
\end{itemize}
In case of the dimers with higher spins, one should consider transitions between
all possible particle-dimer states.

\begin{figure}[t]
\begin{center}
\includegraphics[width=12.cm]{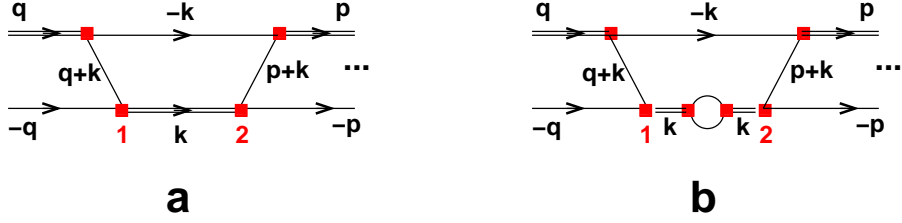}
\caption{Typical diagrams contributing to the particle-dimer scattering. The double line
denotes the dimer, the single line -- the particle, and the filled boxes stand for the
dimer-two-particle vertex.}
\label{fig:dimer_dressed}
\end{center}
\end{figure}

In order to demonstrate, how the expression $p^*\cot\delta(p^*)$ emerges, let us 
consider few diagrams shown in Fig.~\ref{fig:dimer_dressed}. In particular,
in the diagram from Fig.~\ref{fig:dimer_dressed}a, the energy denominators emerging
from the ``hopping'' of a particle between a particle and a dimer are:
\eq
\frac{1}{{\bf k}^2+{\bf q}^2+{\bf k}{\bf q}-mE-i0}
\quad\quad\mbox{and}\quad\quad
\frac{1}{{\bf k}^2+{\bf p}^2+{\bf k}{\bf p}-mE-i0}\, ,
\en
whereas the factors in the dimer-two-particle vertices (marked as 1 and 2 in the figure)
are, respectively
\eq
f_0+f_1\biggl({\bf q}+\frac{\bf k}{2}\biggr)^2+\cdots 
\quad\quad\mbox{and}\quad\quad
f_0+f_1\biggl({\bf p}+\frac{\bf k}{2}\biggr)^2+\cdots\, .
\en
Writing
\eq
 f_0+f_1\biggl({\bf q}+\frac{\bf k}{2}\biggr)^2+\cdots
&=& ({\bf k}^2+{\bf q}^2+{\bf k}{\bf q}-mE)+(f_0+f_1(k^*)^2)+\cdots\, ,
\nonumber\\[2mm]
 f_0+f_1\biggl({\bf p}+\frac{\bf k}{2}\biggr)^2+\cdots
&=& ({\bf k}^2+{\bf p}^2+{\bf k}{\bf p}-mE)+(f_0+f_1(k^*)^2)+\cdots\, ,
\en
where $(k^*)^2=mE-\frac{3}{4}\,{\bf k}^2$, it is immediately seen that the first term
in the above expressions cancels with the pertinent energy denominators. The resulting expression has exactly the form
obtained by inserting the local 
particle-dimer coupling in the diagram. Consequently, it can be removed by redefining the couplings $h_0,h_2,\ldots$.
The remaining piece contains the factor $f^2((k^*)^2)$, 
where $f((k^*)^2)=f_0+f_1(k^*)^2+\cdots$. Moreover, the same factor  $f((k^*)^2)$
emerges from the diagram  Fig.~\ref{fig:dimer_dressed}b as well -- here one may use
dimensional regularization to regulate the loop, so the remainder leads to the no-scale integrals and explicitly vanishes
(albeit the result, of course, does not depend on the regularization). To summarize, in the case without derivative interactions,
the dimer propagator was given by
\eq
\frac{f_0^2}{\sigma}+\frac{f_0^4}{\sigma^2}\,I(k^*)+\cdots
=\frac{1}{\sigma f_0^{-2}-I(k^*)}\propto \frac{1}{-a^{-1}-ik^*}\, ,
\en
where $I(k^*)\propto ik^*$ denotes the loop integral in Fig.~\ref{fig:dimer_dressed}b
with no derivative vertices. If there are derivative vertices present, the above series is modified:
\eq
\frac{f^2((k^*)^2)}{\sigma}+\frac{f^4((k^*)^2)}{\sigma^2}\,I(k^*)+\cdots
=\frac{1}{\sigma f^{-2}((k^*)^2)-I(k^*)}\propto \frac{1}{k^*\cot\delta(k^*)-ik^*}\, .
\en
An important remark is in order. Usually, in the calculations, the quantity $p\cot\delta(p)$
is approximated by the first few terms in the effective-range expansion.
This may lead to the emergence of the spurious poles in the dimer propagator, which
lie far outside the region of applicability of the effective-range expansion, and
which may render the numerical solution of the scattering equation unstable. 
In order to circumvent this problem, e.g.,
in Ref.~\cite{Bedaque:1998km,Hammer:2001gh,Ji:2011qg,Ji:2012nj} a perturbative
expansion of the dimer propagator in the effective radius has been proposed.
A similar problem might also arise in a finite-volume case, which is our primary
concern in this paper. Here we assume that this problem has already been
addressed and solved in the infinite volume -- e.g., by finding a proper parameterization
for  $p\cot\delta(p)$ which, at small values of $p$, reproduces the effective-range 
expansion to a given order, but has a reasonable behavior at large $p^2$
and does not lead to spurious poles in the dimer propagator.

\subsection{Short summary: three-body problem in the infinite volume}

The main features of the three-body problem in the infinite volume can be summarized as follows:

\begin{itemize}

\item[(i)]
All physical observables in the three-particle sector at low energies are parameterized
in terms of the two-particle $C_0,C_2,\ldots$ and three-particle $D_0,D_2,\ldots$ couplings.
Using equations of motion for reducing number of the independent couplings does not
change the observables. The usual power counting applies in the two- and 
three-particle sector.
\item[(ii)]
The particle-dimer picture is not an approximation but an equivalent language for the
description of the three-particle dynamics. In this picture, one trades the
couplings $D_0,D_2$ for the couplings $h_0,h_2,\ldots$.
In order to incorporate the higher partial waves, dimers with arbitrary (integer)
spin should be incorporated.
\item[(iii)]
The three-body scattering $T$-matrix, which contains information about all physical observables in the three-particle sector, can be expressed in terms of the particle-dimer scattering amplitude ${\cal M}$, which obeys the equation~(\ref{eq:BSA}), modified in the presence of derivative couplings. Note that the quantity $\tau({\bf k};E)$ in this equation is already parameterized in terms of the physical observables (phase shift) and does not
explicitly depend on the regularization used.
 
\end{itemize}

\section{Finite volume}
\label{sec:finite}

\subsection{Strategy}

In the previous section, we have thoroughly considered the formulation of the 
three-particle problem in the infinite volume and have demonstrated that the 
particle-dimer picture provides an equivalent description of this problem.
The reason why we have done this is simple -- we will show that the quantization
condition can be obtained almost immediately and in an absolutely transparent manner,
assuming that the three-momenta in a finite volume are quantized.

The main question here is the choice of an appropriate
set of physical observables (parameters), which should be determined from
the measured finite volume spectrum. As mentioned before, the previous attempts
in the literature
were focused on splitting the equations in a finite volume into the infinite-volume
part and the rest. We believe that the most convenient choice of the parameters
is provided by the low-energy couplings themselves -- once these couplings are determined on the
lattice, the particle-dimer scattering amplitude 
can be constructed by solving the scattering equation in the infinite volume.

Let us explain the procedure in more detail. The ``scattering amplitude''
 in a finite volume is determined by the equation\footnote{Here we restrict ourselves again to the case of a scalar dimer only. Higher partial waves can be included straightforwardly in a later stage.}
\eq\label{eq:BSAL}
{\cal M}_L({\bf p},{\bf q};E)=Z({\bf p},{\bf q};E)
+\frac{8\pi}{L^3}\sum\limits_{\bf k}^\Lambda\,Z({\bf p},{\bf k};E)\tau_L({\bf k};E)
{\cal M}_L({\bf k},{\bf q};E)\, ,
\en
where ${\bf k}=\frac{2\pi}{L}\,{\bf n}\, ,~{\bf n}\in \mathbb{Z}^3$ is a quantized three-momentum in a finite volume and
\eq\label{eq:tauL}
\tau^{-1}_L({\bf k};E)=k^*\cot\delta(k^*)+S({\bf k},(k^*)^2)\, ,
\en
where
\eq\label{eq:S}
S({\bf k},(k^*)^2)=
-\frac{4\pi}{L^3}\sum_{\bf l}\frac{1}{{\bf k}^2+{\bf l}^2+{\bf k}{\bf l}-mE}\, .
\en
This sum diverges in the ultraviolet and needs to be regularized and renormalized.
The simplest way to do this is to perform a subtraction at some $(k^*)^2=-\mu^2<0$
(one subtraction suffices, but more subtractions lead to a faster convergence).
Then, we get
\eq
S({\bf k},(k^*)^2)=\bigl[S({\bf k},(k^*)^2)-S({\bf k},-\mu^2)\bigr]
+S({\bf k},-\mu^2)\, .
\en
The first term in brackets is finite. The second term is equal to
\eq
S({\bf k},-\mu^2)
=-\frac{4\pi}{L^3}\sum_{\bf l}\frac{1}{\bigl({\bf l}+{\bf k}/2\bigr)^2+\mu^2}
=\mu-\sum_{{\bf n}\neq{\bf 0}}\frac{1}{nL}\,\exp\biggl(-\frac{i}{2}\,L{\bf k}{\bf n}
-nL\mu\biggr)\, ,
\en
where $n=|{\bf n}|$ and we have used Poisson's summation formula. Note that the ultraviolet divergence in the above expression can be tamed, e.g., by using dimensional regularization. 
Note also that the equation~(\ref{eq:BSAL}) was used earlier
in Refs.~\cite{Kreuzer:2010ti,Kreuzer:2009jp,Kreuzer:2008bi,Kreuzer:2012sr} to
calculate the spectrum of the three-body bound states in a finite volume numerically.

Now, our strategy for the analysis of data in the three-particle sector can be formulated
as follows:

\begin{enumerate}

\item
Consider first the two-particle sector, extract the phase shift $\delta(p)$ at different
momenta by using the L\"uscher equation. Parameterize the function $p\cot\delta(p)$
so that it fits the lattice data and does not lead to spurious poles at large momenta.

\item
Fix the cutoff $\Lambda$.
Truncate the partial-wave expansion (consider dimers with a spin below some fixed value).
Fit the spectrum in the three-particle sector, using $h_0,h_2,\ldots$ as free parameters.
Repeat this until the fit does not improve anymore by adding parameters.

\item
Solve the equations in the infinite volume by using the same values of the parameters
and the same cutoff $\Lambda$. Calculate different cross sections, bound-state
energies, etc.

\end{enumerate}

The proposed scheme has apparent advantages with respect to the ones proposed
in the literature. First of all, it is extremely simple. For example, since we do not
want to single out the scattering amplitude in the infinite volume, we do not need to
introduce a ``smooth cutoff'' on the momentum ${\bf l}$ in Eq.~(\ref{eq:S}).
Furthermore, the procedure is systematic: the particle-dimer coupling constants
$h_0,h_2,\ldots$, corresponding to a dimer with a fixed spin,
obey the usual counting rules. At the lowest order in the momentum expansion it is 
just one constant $h_0$ for the scalar dimer, which describes the whole three-body
spectrum. Finally, the minimal set of couplings is observable, in the sense that
they can be uniquely determined from matching to the on-shell $T$-matrix.

In view of the above discussion, the fundamental 
statement that the finite-volume energy 
spectrum is determined solely by the three-body $S$-matrix, can be rephrased as
follows: the off-shell couplings (like $D_4''$ considered in the previous section)
have no impact on the finite-volume spectrum. This statement can be checked
immediately by using exactly the same diagrammatic technique as in the previous
section. Indeed, using the no-scale argument, we have proven that $D_4''$ does not 
enter the on-shell three-body $T$-matrix. The finite-volume counterpart of this
$T$-matrix, which is defined by the same set of Feynman integrals with 3-momentum
integrations replaced by sums, determines the spectrum of the three-body system in a 
finite volume. Namely, its poles correspond to the energy levels. The no-scale integrals
in a finite volume also give vanishing contribution (to be more precise, the contribution
from the off-shell LECs will be exponentially suppressed $\propto\exp(-ML)$, where
the scale $M\sim m$ is much larger than typical non-relativistic momenta). Thus,
the fundamental statement from Refs.~\cite{Polejaeva:2012ut,Hansen:2014eka}
looks almost trivial in the new framework. 

\subsection{Application}

In order to demonstrate, how the proposed framework works in practice, we have done
a simple exercise. We have restricted ourselves to the scalar dimer and neglected
derivative couplings, both in the two-particle and in the three-particle sectors. Thus,
for a fixed parameter $\Lambda$, we have two free parameters -- the two-body
scattering length $a$ and the particle-dimer coupling $h_0(\Lambda)$ which
can be traded for $H_0(\Lambda)$, see subsection~\ref{sec:scatt_eq}. Hence, we want to find
the poles of the amplitude ${\cal M}_L$, defined by the Eq.~(\ref{eq:BSAL}),
where Eq.~(\ref{eq:tauL}) takes the form
\eq\label{eq:tauLn}
\tau^{-1}_L({\bf k};E)=-a^{-1}+S({\bf k},(k^*)^2)\, ,
\en
and
\eq
Z({\bf p},{\bf q};E)=\frac{1}{-mE+{\bf p}^2+{\bf q}^2+{\bf p}{\bf q}}
+\frac{H_0(\Lambda)}{\Lambda^2}\, .
\en
In the vicinity of a pole $E=E_n$, the amplitude ${\cal M}_L$ factorizes
\eq
{\cal M}_L({\bf p},{\bf q};E)=\frac{{\cal F}({\bf p}){\cal F}({\bf q})}{E-E_n}+\mbox{regular terms.}
\en
Consequently, the spectrum will be determined from the following homogeneous equation:
\eq\label{eq:calF}
{\cal F}({\bf p})=\frac{8\pi}{L^3}\,\sum_{\bf q}^\Lambda
Z({\bf p},{\bf q};E)\tau_L({\bf q};E){\cal F}({\bf q})\, .
\en
Next, we would like to perform a partial-wave expansion in this equation. Note that,
since the rotation symmetry is broken down by the cubic lattice, a mixing of the partial
waves will occur. Using the so-called cubic harmonics, the maximal
diagonalization of the matrix equation can be achieved. In doing so, we shall use the
general formalism described in Refs.~\cite{Bernard:2008ax,Gockeler:2012yj}
The final result coincides with that of Ref.~\cite{Kreuzer:2012sr}.

We start with the expanding of $Z$ and ${\cal F}$ in partial waves
\eq
Z({\bf p},{\bf q};E)&=&4\pi\sum_{lm}Y_{lm}(\hat p)Z_l(p,q;E)Y^*_{lm}(\hat q)\, ,
\nonumber\\[2mm]
{\cal F}({\bf p})&=&\sqrt{4\pi}\sum_{lm}Y_{lm}(\hat p){\cal F}_{lm}(p)\, ,
\en
where $\hat p,\hat q$ are unit vectors in the direction of ${\bf p},{\bf q}$, and
$Y_{lm}$ denotes spherical functions. Since the spherical symmetry is broken,
${\cal F}_{lm}$ depends on $m$. Further,
\eq
Z_l(p,q;E)=-\frac{1}{2pq}\,\mbox{Re}\,Q_l\biggl(\frac{mE-p^2-q^2}{pq}\biggr)+\frac{H_0(\Lambda)}{\Lambda^2}\,\delta_{l0}\, ,
\en
where $Q_l(z)=\frac{1}{2}\,\int_{-1}^1dx\frac{P_l(x)}{z-x}$ is the Legendre function of the second kind. In particular,
\eq
Z_0(p,q;E)=\frac{1}{2pq}\,\ln\biggl|\frac{p^2+pq+q^2-mE}{p^2-pq+q^2-mE}\biggr|+\frac{H_0(\Lambda)}{\Lambda^2}\, .
\en
The equation (\ref{eq:calF}) can be now rewritten in the form
\eq\label{eq:lindet}
{\cal F}_{lm}(p)=\sum_q^\Lambda\sum_{l'm'}Z_l(p,q;E)R_{lm,l'm'}(q;E){\cal F}_{l'm'}(q)\, ,
\en
where
\eq
R_{lm,l'm'}(q;E)=\frac{8\pi}{L^3}\,\sum_{|{\bf q}|=q}
Y^*_{lm}(\hat q)\tau_L({\bf q};E)Y_{l'm'}(\hat q)\, ,
\en
and $p,q$ take the discrete values:
$p,q=\frac{2\pi}{L}\,\sqrt{n_1^2+n_2^2+n_3^2}\,,~n_1,n_2,n_3\in\mathbb{Z}$. 

\begin{figure}[t]
\begin{center}
\includegraphics*[width=10.cm]{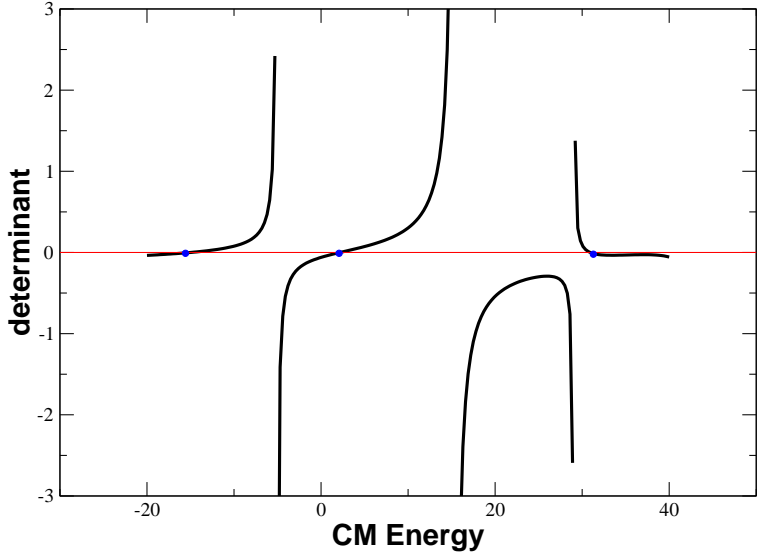}
\caption{The determinant from Eq.~(\ref{eq:quant_sym}). The values of the parameters
are $\Lambda a=225$ $L/a=1$, while the shaded blobs correspond to the position
of the energy levels.}
\label{fig:determinant}
\end{center}
\end{figure}

The system of the homogeneous linear equations~(\ref{eq:lindet}) has a solution if and
only if its determinant vanishes. Consequently, the quantization condition takes the form
\eq\label{eq:quant_lm}
\det\biggl(\delta_{ll'}\delta_{mm'}\delta_{pq}-Z_l(p,q;E)R_{lm,l'm'}(q;E)\biggr)=0\, .
\en
We see explicitly that partial-wave mixing occurs in the quantization condition
due to the lost rotational symmetry. It is still possible to block-diagonalize this equation,
using the remnant cubic symmetry on the lattice. The basis vectors of the irreducible
representations of the cubic group are given by the linear combinations of those
of the rotation group~\cite{Bernard:2008ax,Gockeler:2012yj}
\eq
|\Gamma\alpha ln\rangle=\sum_m c^{\Gamma\alpha n}_{lm}|lm\rangle\, .
\en
Here, $\Gamma$ denotes one of the five irreducible representations $A_1,A_2,E,T_1,T_2$
of the cubic group, $n=1,\ldots N(\Gamma,l)$, where $N(\Gamma,l)$ is the number of
occurrences of $\Gamma$ in the irreducible representation $D^l$ of the rotation group,
and $\alpha=1,\ldots \mbox{dim}(\Gamma)$ labels the basis vectors in the irreducible
representation $\Gamma$. Further, $c^{\Gamma\alpha n}_{lm}$ are Clebsch-Gordan
coefficients. Next, we define
\eq
R^{\Gamma\Gamma'}_{ln\alpha,l'n'\alpha'}(q;E)=\sum_{mm'}
(c^{\Gamma\alpha n}_{lm})^*R_{lm,l'm'}(q;E) c^{\Gamma'\alpha' n'}_{l'm'}
=\delta_{\Gamma\Gamma'}\delta_{\alpha\alpha'}R^{\Gamma}_{ln,l'n'}(q;E)\, ,
\en
where the last equality was obtained by using Schur's lemma. Using the orthogonality
of the Clebsch-Gordan coefficients, Eq.~(\ref{eq:quant_lm}) can be rewritten as follows:
\eq
\det\biggl(\delta_{nn'}\delta_{ll'}\delta_{pq}-Z_l(p,q;E)R^\Gamma_{ln,l'n'}(q;E)\biggr)=0\, .
\en
For simplicity, let us restrict ourselves to the S-waves $l=l'=0$. Then, only 
$\Gamma=A_1$ appears in the above expansion and we get
\eq
\det\biggl(\delta_{pq}-Z_0(p,q;E)R^{A_1}(q;E)\biggr)=0\, ,
\en
where
\eq
R^{A_1}(q)=\frac{2}{L^3}\,\sum_{|{\bf q}|=q}\tau_L({\bf q};E)\, .
\en
The matrix can be symmetrized, leading to the eigenvalue equation
\eq\label{eq:quant_sym}
\det\biggl(s(p;E)\delta_{pq}-\sqrt{|R^{A_1}(p;E)|}Z_0(p,q;E)\sqrt{|R^{A_1}(q;E)|}\biggr)=0\, ,
\en
where $s(p;E)=R^{A_1}(q;E)/|R^{A_1}(q;E)|$.

\begin{figure}[t]
\begin{center}
\includegraphics*[width=10.cm]{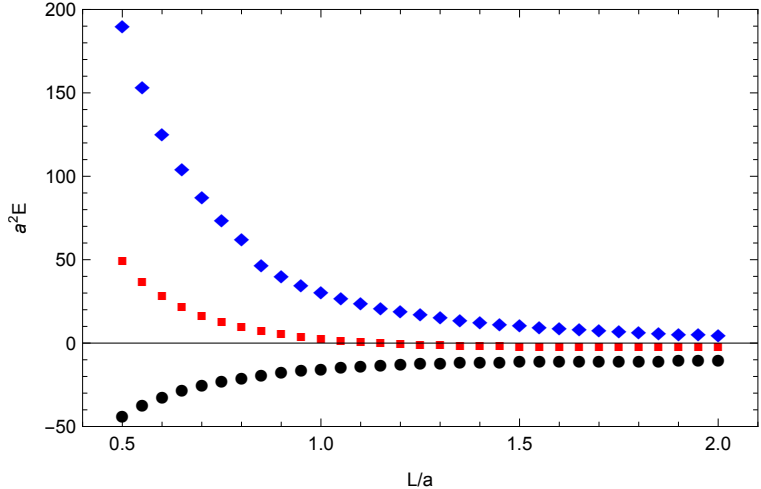}
\caption{The dependence of the energy levels on the box size for a fixed value
$\Lambda a=225$. The levels both below and above the three-particle threshold
are considered. The behavior of the subthreshold level agrees well with the result of
Ref.~\cite{Kreuzer:2012sr}.}
\label{fig:spectrum}
\end{center}
\end{figure}

The equation~(\ref{eq:quant_sym}) can be solved numerically. The square matrix is 
finite, since $p,q$ are both bound by $\Lambda$. 
We choose the parameters as in Ref.~\cite{Kreuzer:2012sr}. First of all, we take $m=1$.
Further, the cutoff $\Lambda$ is fixed through the two-body scattering length, $a$, by
$\Lambda a=225$. Finally, we require that the energy of the bound state in the 
 infinite volume is equal to $E_\infty=-10(ma^2)^{-1}$. This fixes the coupling constant $H_0=0.1925$.
Calculating now the determinant in Eq.~(\ref{eq:quant_sym}) as a 
function of $E$, one may find the discrete values of $E=E_1,E_2,\ldots$, where
this function crosses the horizontal axis. Repeating the calculations at different
values of the box size $L$, one gets the volume-dependent spectrum $E_1(L),E_2(L),\ldots$.
Figure~\ref{fig:determinant} shows the results of calculation of the determinant at
one particular value of $L$, whereas Fig.~\ref{fig:spectrum} displays the volume-dependent spectrum
-- both below and above three-particle threshold.

What does one learn from these results? For a fixed value of $\Lambda$ and $L$,
the spectrum is determined by a limited set of the parameters $H_0,H_2,\ldots$
(we imply that $C_0,C_2,\ldots$ are determined in the two-particle sector). In the simplest
case considered here, this is just one parameter $H_0$. For the analysis of the lattice data,
one has to calculate the energy levels as a function of $H_0$ and fit to the existing data.
If this is not enough, one has to include derivative couplings and dimers with higher spin,
until the quality of fit does not change. We stress once more that this is a systematic prescription
-- the derivative couplings obey counting rules and are less and less relevant at low energies.
The same is true for the partial-wave truncation -- higher partial ways give little contribution at low energies.

To summarize, the framework described above gives a simple and systematic tool
to analyze the lattice data in the three-particle sector. 

\section{Comparison with existing approaches}
\label{sec:comparison}

Several groups have previously addressed the three-particle problem in a finite volume.
In this section, we would like to briefly review these approaches and clarify the
relation to the approach which was considered in the present paper.

We start from Refs.~\cite{Hansen:2014eka,Hansen:2015zga,Briceno:2017tce}, where
the three-particle quantization condition has been obtained. That derivation has been carried out
looking for the poles of the three-particle Green function in a finite volume.
Since we have explicitly demonstrated that the particle-dimer picture is equivalent to the
three-particle description, and since our formalism also extracts the poles of the 
particle-dimer scattering amplitude, it is no wonder that, {\em diagrammatically,}
both quantization conditions are the same. This will be explicitly shown below.

\begin{figure}[t]
\begin{center}
\includegraphics[width=10.cm]{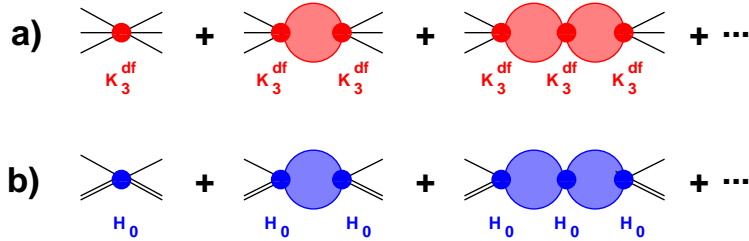}
\caption{a) The three-particle Green function and b) The particle-dimer Green function:
insertion of the three particle/particle-dimer couplings. The shaded blobs contain only
self-energy insertions and particle ``hopping,'' see Fig.~\ref{fig:blobs}.}
\label{fig:comp-Hansen}
\end{center}
\end{figure}

\begin{figure}[t]
\begin{center}
\includegraphics[width=14.cm]{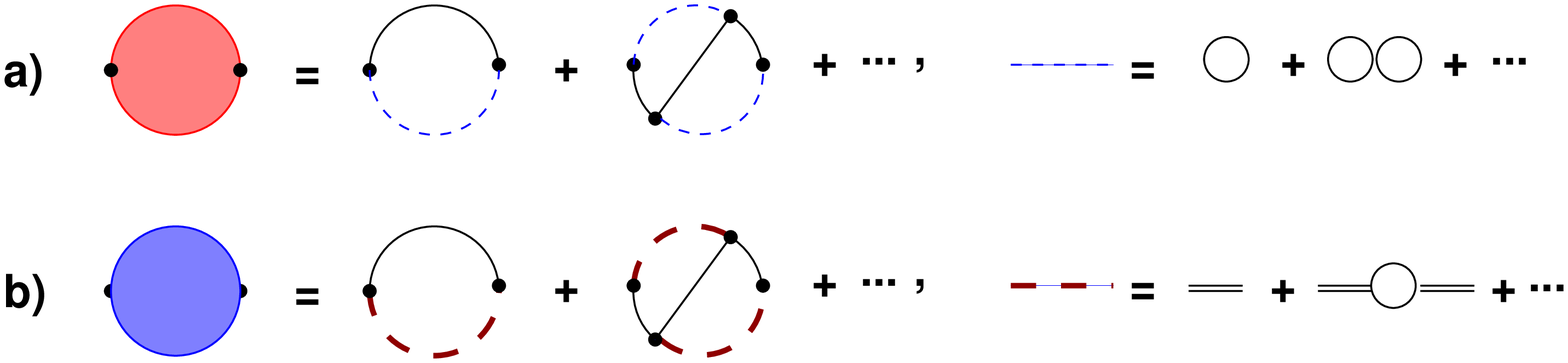}
\caption{The shaded blobs in Fig.~\ref{fig:comp-Hansen} describe the propagation of
the spectator particle and a) a pair of particles interacting with each other through
contact interactions, or b) a dressed dimer. These two lines are connected by particle
``hopping.''}
\label{fig:blobs}
\end{center}
\end{figure}

First, note that the quantity $K_3^{\sf df}$, defined in Ref.~\cite{Hansen:2014eka}, is a counterpart of our
$H_0+\ldots$ (a more precise statement will follow). 
The three-particle Green function in the framework of Refs.~\cite{Hansen:2014eka,Hansen:2015zga,Briceno:2017tce} and
the particle-dimer Green function in our approach
are shown in Figs.~\ref{fig:comp-Hansen}a and \ref{fig:comp-Hansen}b, respectively.
It is clear that one may relate these two quantization conditions, if the corresponding shaded blobs
(containing only pair interactions), or only two-particle dimer interactions, are related to each other.
These blobs are depicted in Figs.~\ref{fig:blobs}a and \ref{fig:blobs}b, respectively.

\begin{figure}[t]
\begin{center}
\includegraphics[width=12.cm]{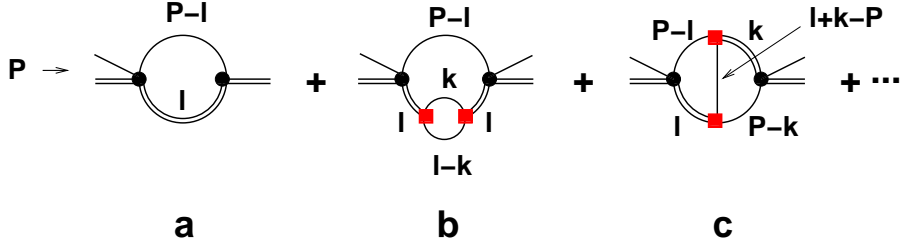}
\caption{The lowest-order diagrams, contributing to the shaded blob in the
particle-dimer picture, see Fig.~\ref{fig:blobs}b. The double line depicts a dimer and the single line -- a particle.}
\label{fig:ourblob}
\end{center}
\end{figure}

The blob in Fig.\ref{fig:blobs}b contains two-types of diagrams: self-energy insertions
into the dimer line, and the particle ``hopping.'' In order to get a feeling, let us calculate
few diagrams shown in Fig.~\ref{fig:ourblob} (for simplicity, let us restrict ourselves to the non-derivative
couplings only).
Using split dimensional regularization in the first term\footnote{Using split dimensional
regularization boils down to declaring the $l_0$-integral
in the first term equal to zero, since the pole is on one side of the axis. Note that
this prescription {\em was not used} in the derivation of Eqs.~(\ref{eq:BSA}) and (\ref{eq:BSAL}). However,
it can be shown that the change of the regularization is equivalent to the redefinition of the coupling $H_0$ and
is thus physically irrelevant.}, it can be shown that
\eq\label{eq:blob}
{\cal M}_a&=&\intsuml \frac{1}{\sigma}\,\frac{1}{\frac{{\bf l}^2}{2m}-E+l_0-i0}=0\, ,
\nonumber\\[2mm]
{\cal M}_b&=&\frac{1}{2}\,f_0^2\intsuml\intsumk\frac{1}{\sigma^2}
\frac{1}{\frac{{\bf l}^2}{2m}-E+l_0-i0}\,\frac{1}{\frac{{\bf k}^2}{2m}-k_0-i0}\,
\frac{1}{\frac{({\bf l}+{\bf k})^2}{2m}-l_0+k_0-i0}
\nonumber\\[2mm]
&=&\frac{1}{2}\,f_0^2\frac{1}{L^3}\sum_{\bf l}\frac{1}{L^3}\sum_{\bf k}
\frac{1}{\frac{1}{m}\,({\bf l}^2+{\bf k}^2+{\bf l}{\bf k})-E}\, ,
\nonumber\\[2mm]
{\cal M}_c&=&f_0^2\intsuml\intsumk\frac{1}{\sigma^2}
\frac{1}{\frac{{\bf l}^2}{2m}-P_0+l_0-i0}\,\frac{1}{\frac{{\bf k}^2}{2m}-P_0+k_0-i0}\,
\frac{1}{\frac{({\bf l}+{\bf k})^2}{2m}-P_0-k_0+l_0-i0}
\nonumber\\[2mm]
&=&f_0^2\frac{1}{L^3}\sum_{\bf l}\frac{1}{L^3}\sum_{\bf k}\frac{1}{\frac{1}{m}\,
({\bf l}^2+{\bf k}^2+{\bf l}{\bf k})-E}\, ,
\en
and so on, where
\eq
\intsuml(\cdots)=\int\frac{dl_0}{2\pi i}\frac{1}{L^3}\sum_{\bf l}(\cdots)\, .
\en
Denoting the self-energy insertion by $J$ and the ``hopping'' diagram by $-G$, and defining $a_0=\frac{1}{2}\,f_0^2$,
the blob shown in Fig.~\ref{fig:blobs}b can be symbolically written
down as
\eq\label{eq:Mdimer}
{\cal M}_{dimer}&=&(a_0^2J+a_0^3J^2+\cdots)
-(a_0+a_0^2J+\cdots)G(a_0+a_0^2J+\cdots)+\cdots
\nonumber\\[2mm]
&=&\frac{a_0^2J}{1-a_0J}-\frac{a_0J}{1-a_0J}\,G\frac{a_0J}{1-a_0J}+\cdots\, .
\en
Next, we prove a key equality that allows one to transform this expression into
the form given in Refs.~\cite{Hansen:2014eka,Hansen:2015zga,Briceno:2017tce}.
It is seen from Eq.~(\ref{eq:blob}) that, if $G$ appears on the left or on
the right of all insertions, then $G$ can be replaced by $-2J$. This is a general
property which is valid for any number of insertions. Indeed, as seen from Fig.~\ref{fig:ourblob},
contracting the free dimer propagator to a point, topologically there is no
difference between the diagrams Fig.~\ref{fig:ourblob}b and Fig.~\ref{fig:ourblob}c.
This will remain true, if any number of insertions (both $J$ and $G$) are placed on
the left or on the right (but not both!).

Now, we are able to demonstrate that our quantization condition is equivalent to that
of Refs.~\cite{Hansen:2014eka,Hansen:2015zga,Briceno:2017tce}. To this end, we
introduce the following notations for the two- and three-body amplitudes:
\eq\label{eq:M23}
{\cal M}_2=\frac{a_0}{1-a_0J}\, ,\quad\quad
{\cal M}_3={\cal M}_2-{\cal M}_2G{\cal M}_3\, .
\en
Using these definitions and replacing $G$ by $(-2J)$ on the left and on the right in all
terms, after lengthy but straightforward algebraic transformations 
(see appendix~\ref{app:Hansen-Sharpe}), we obtain
\eq\label{eq:1over3}
{\cal M}_{dimer}=9a_0^2\biggl(\frac{1}{3}\,J+J{\cal M}_3J\biggr)\, .
\en
This is exactly the same expression as in Refs.~\cite{Hansen:2014eka,Hansen:2015zga,Briceno:2017tce}.

Though we have just demonstrated that {\em diagrammatically} both quantization
conditions are equivalent, some differences remain. In the Refs.~\cite{Hansen:2014eka,Hansen:2015zga,Briceno:2017tce}
a smooth cutoff, introduced first in Ref.~\cite{Polejaeva:2012ut}, has been used both in the
self-energy insertion $J$ as well as on the
sum over the momenta ${\bf q}$ of the spectator particle. Denoting the cutoff function
by $H({\bf q})$, the sum over ${\bf q}$ is given by
\eq
\frac{1}{L^3}\sum_{\bf q}(\cdots)=\frac{1}{L^3}\sum_{\bf q}\bigl[H({\bf q})
+(1-H({\bf q}))\bigr](\cdots)\, .
\en
The cutoff function is chosen so that $(1-H({\bf q}))=0$, if $mE>\frac{3}{4}\,{\bf q}^2$
(in the non-relativistic limit). This means that the second term in this equation is smooth and
the sum can be replaced by the integral. Consider now Eq.~(\ref{eq:BSAL}). Symbolically, it can be written as
\eq
{\cal M}_L=Z+\sum_{\bf q}H({\bf q})Z\tau_L {\cal M}_L+\int_{\bf q}(1-H({\bf q}))
Z\tau {\cal M}_L\, ,
\en
where $\tau=\lim_{L\to\infty}\tau_L$.
Defining now the ``non-standard $K$-matrix'' in the context of the particle-dimer picture
\eq
{\cal M}_H=Z+\int_{\bf q}(1-H({\bf q}))Z\tau {\cal M}_H\, ,
\en
we arrive at
\eq
{\cal M}_L={\cal M}_H
+\sum_{\bf q}H({\bf q}){\cal M}_H\tau_L {\cal M}_L\, .
\en
The quantization condition that can be obtained from this equation has 
exactly the same form as in Refs.~\cite{Hansen:2014eka,Hansen:2015zga,Briceno:2017tce}.
Consequently, the main difference with our approach consists in moving down the cutoff from $\Lambda$.
The advantage of this is that the system
of the linear equations that determines the finite-volume spectrum is lower-dimensional,
since the maximum momentum is determined by the cutoff. However, one has to introduce the quantity ${\cal M}_H$ instead
of a single parameter $H_0$, and further find
the conventional infinite-volume $K$ matrix by solving an integral equation involving
${\cal M}_H$. All this renders the formalism complicated and difficult to apply
to the analysis of data.

In addition, Ref.~\cite{Hansen:2014eka} 
it was argued that the introduction of the cutoff $H$ allows one to always define
the Lorentz boost, bringing the pair of two interacting particles to the rest frame.
Namely, the cutoff excludes the region, where the total momentum squared of two 
particles is negative. While the latter statement is certainly true, in this case
the problem is merely shifted to the calculation of ${\cal M}_H$. Note also that
the same problem would arise in the three-body problem in the infinite volume
as well, leading to the conclusion that one can not move the cutoff above some fixed
point determined by kinematics. The fact is that, even if the boost is undefined, the
two-body scattering amplitude is defined for all momenta, given the non-relativistic 
Lagrangian, in which the couplings are matched to the effective-range expansion
parameters defined in the CM frame. In the relativistic case, it is convenient
to use the covariant version of the non-relativistic effective field 
theory~\cite{Colangelo:2006va,Gasser:2011ju}, which provides explicitly Lorentz-invariant
expression of the two-body amplitudes.  

Having considered Refs.~\cite{Hansen:2014eka,Hansen:2015zga,Briceno:2017tce} in
detail, we next turn to Ref.~\cite{Polejaeva:2012ut}. The three-particle Green function,
considered here, contains the same set of diagrams as in Refs.~\cite{Hansen:2014eka,Hansen:2015zga,Briceno:2017tce} (including, in particular, two-to-three transitions as well). The final answer is given in a form of a system of linear equations, and the quantization condition can straightforwardly be obtained by declaring the determinant of this system equal to zero.
What is different, is the cutoff on the spectator momentum, which is now moved down to zero (the cutoff in the self-energy diagram stays where it was before). Since L\"uscher's 
regular summation theorem in the spectator momentum ${\bf q}$
can not be applied everywhere, the dimensionality of the equations is the same
as in Refs.~\cite{Hansen:2014eka,Hansen:2015zga,Briceno:2017tce} and, because
the ``non-standard $K$-matrix'' has to be introduced, this formalism is very complicated
as well.

Finally, the formalism constructed in Ref.~\cite{Briceno:2012rv} is based on the
particle-dimer picture and is close to the one we are using here. The authors go further
and replace the exact dimer propagator by a sum over the finite-volume two-particle
poles (there is always a finite number thereof, limited by kinematics) and the rest, that
can be calculated in the infinite volume. This effectively amounts to introducing a cutoff
on the spectator momentum, defined by the lowest-lying pole. All above discussion also applies here.

To summarize, the alternative approaches discussed in this section
separate finite- and infinite-volume contributions in the three-particle amplitude.
This leads to the introduction of the quantities like the ``unconventional $K$-matrix''
that render the formalism rather complicated. Instead of this, we advocate 
for solving the particle-dimer equation in a finite volume directly and fitting the 
low-energy couplings to the measured spectrum. Note also that, since all these
unconventional $K$-matrices should be low-energy polynomials (the low-energy
region is cut off in the integral equation), one could expand these quantities in momenta
and arrive at equations which are formally similar to our equations but are using
a smooth cutoff at $\Lambda\sim m$. Exactly the same result can be obtained in much more direct way,
lowering the cutoff  $\Lambda$ in our equations and taking into account the fact that the couplings
run with $\Lambda$.

\section{Conclusions}
\label{sec:concl}

In this paper, we propose an effective field theory framework, which can be
used to analyze the three-particle spectrum on the lattice.

\begin{itemize}
\item[(i)]
The crucial point consists
in the choice of a convenient set of parameters that can be fit to the lattice data.
Within our framework, this set is formed by the particle-dimer couplings
(effective couplings in the two-particle sector should be determined separately).
Any physical observable can be determined, using the set of couplings determined
on the lattice.  

\item[(ii)]
The proposed approach is hardly new. However, it was important to realize that it
is algebraically equivalent to existing approaches, known in the literature.
Moreover, we have demonstrated that it is systematic, 
much simpler in use than the existing approaches, and can be applied
for data analysis even when only a few data points are available.

\item[(iii)]
As an illustration, we have used this approach to calculate the energy spectrum (both 
below and above the three-particle threshold) by using the input values of the couplings.
As a further illustration, note that this approach was successfully used~\cite{1}
 to reproduce the
result of Ref.~\cite{Meissner:2014dea} on the three-body bound state energy shift
in a finite volume, as well as to study the role of the three-particle force and the
generalization of the above result beyond the unitary limit.

\item[(iv)]
In order to emphasize the main conceptual problem and its solution,
technical details like relativistic kinematics, higher partial waves, etc.
have been left out. These will be included in future publications.

\item[(v)]
The treatment in this paper rests on a particular form of the effective field theory
that describes the three-particle system in a finite volume. In the case of the nucleons, for
example, one may have to take into account the pion exchanges between nucleons explicitly
leading to a different ``chiral'' effective field theory. This does not, however, change
the underlying idea of our proposal -- considering a given effective field theory
(relativistic or non-relativistic) in a finite volume, calculating the spectrum numerically and fitting
the low-energy constants to the spectrum. Note also that a very similar strategy
is pursued within the framework of the unitary ChPT in a finite
volume~\cite{oset,Doring-scalar,oset-in-a-finite-volume}, which is used in the analysis of data in the
two-particle sector. A possible alternative scheme in the three-body sector based on the isobar approximation
was recently discussed in Ref.~\cite{Mai:2017vot}.

\end{itemize}

\begin{acknowledgments}

The authors would like to thank R. Briceno, Z. Davoudi, M. D\"oring, 
E. Epelbaum, M. Hansen, D. Lee, T. Luu, 
M. Mai, U.-G. Mei{\ss}ner and S. Sharpe for useful discussions.  

We acknowledge support from the DFG through funds provided to the
Sino-German CRC 110 ``Symmetries and the Emergence of Structure in QCD'' and the
CRC 1245 ``Nuclei: From Fundamental Interactions to Structure and Stars'' as well as
the BMBF under contract 05P15RDFN1.
This research is also supported in part by Volkswagenstiftung
under contract no. 86260 
and by Shota Rustaveli National Science Foundation (SRNSF), grant no. DI-2016-26.

\end{acknowledgments}

\newpage

\appendix

\section{Derivation of Eq.~(\ref{eq:1over3})}
\label{app:Hansen-Sharpe}

The equation~(\ref{eq:1over3}) can be obtained by transforming the expression~(\ref{eq:Mdimer}) as
\eq
a_0^{-2}{\cal M}_{dimer}&=&J\frac{1}{1-a_0J}
-\frac{1}{1-a_0J}\,\biggl\{G-G\frac{a_0}{1-a_0J}G+G\frac{a_0}{1-a_0J}\,G\frac{a_0}{1-a_0J}G-\cdots\biggr\}\frac{1}{1-a_0J}
\nonumber\\[2mm]
&=&J\frac{1}{1-a_0J}-\frac{1}{1-a_0J}\,\biggl\{G-G({\cal M}_2-{\cal M}_2G{\cal M}_2+\cdots)G\biggr\}\frac{1}{1-a_0J}
\nonumber\\[2mm]
&=&J\frac{1}{1-a_0J}-\frac{1}{1-a_0J}\,\biggl\{G-G{\cal M}_3G\biggr\}\frac{1}{1-a_0J}
\nonumber\\[2mm]
&=&J\frac{1}{1-a_0J}-G-\frac{a_0J}{1-a_0J}\,G-G\frac{a_0J}{1-a_0J}
-\frac{a_0J}{1-a_0J}\,G\frac{a_0J}{1-a_0J}
\nonumber\\[2mm]
&+&G{\cal M}_3G
+\frac{a_0J}{1-a_0J}\,G{\cal M}_3G+G{\cal M}_3G\frac{a_0J}{1-a_0J}
+\frac{a_0J}{1-a_0J}\,G{\cal M}_3G\frac{a_0J}{1-a_0J}\, .
\en
In some terms of the above expression, $G$ appears on the very left or on the very right.
Here, one may replace $G$ with $-2J$, arriving at the following expression
\eq
a_0^{-2}{\cal M}_{dimer}&=&J\frac{1}{1-a_0J}+2J+\frac{4a_0J^2}{1-a_0J}
-\frac{a_0J}{1-a_0J}\,G\frac{a_0J}{1-a_0J}
\nonumber\\[2mm]
&+&4J{\cal M}_3J
-\frac{2a_0J}{1-a_0J}\,G{\cal M}_3J
-J{\cal M}_3G\frac{2a_0J}{1-a_0J}
+\frac{a_0J}{1-a_0J}\,G{\cal M}_3G\frac{a_0J}{1-a_0J}
\nonumber\\[2mm]
&=&J\biggl\{{\cal M}_2G{\cal M}_3G{\cal M}_2-{\cal M}_2G{\cal M}_2+4{\cal M}_3
-2{\cal M}_2G{\cal M}_3-2{\cal M}_3G{\cal M}_2\biggr\}J
\nonumber\\[2mm]
&+&J\frac{1}{1-a_0J}+2J+\frac{4a_0J^2}{1-a_0J}\, .
\en
The last term can be replaced by
\eq
\frac{4a_0J^2}{1-a_0J}=4J{\cal M}_2J\, .
\en
Finally, using Eq.~(\ref{eq:M23}), we get
\eq
a_0^{-2}{\cal M}_{dimer}&=&9J{\cal M}_3J+3J-J({\cal M}_2J+1)+\frac{J}{1-a_0J}
=9\biggl(\frac{1}{3}J+J{\cal M}_3J\biggr)\, ,
\en
which coincides with Eq.~(\ref{eq:1over3})\, .

\appendix

\end{document}